\documentclass[11pt]{article}
\usepackage{amssymb}
\begin{document}
\def\nmonth{\ifcase\month\ \or January\or
   February\or March\or
April\or May\or June\or July\or August\or
   September\or October\or
November\else December\fi}
\def\nmonth{\ifcase\month\ \or January\or
   February\or March\or April\or May\or June\or July\or August\or
   September\or October\or November\else December\fi}
\def\rightheadline{\hfill\folio\hfill}
\def\leftheadline{\hfill\folio\hfill}
\newtheorem{theorem}{Theorem}[section]
\newtheorem{lemma}[theorem]{Lemma}
\newtheorem{remark}[theorem]{Remark}
\def\operatorname#1{{\rm#1\,}}
\def\text#1{{\hbox{#1}}}
\def\qedbox{\hbox{$\rlap{$\sqcap$}\sqcup$}}
\def\BB{{\mathcal{B}_1}}
\def\CC{{\mathcal{B}_2}}
\def\BX{{\mathcal{B}_{DR}}}
\def\BD{{\mathcal{B}_D}}
\def\BR{{\mathcal{B}_R}}
\def\B{{\mathcal{B}}}
\def\tr{{\operatorname{Tr}}}
\def\dvol{{\operatorname{dvol}}}
\newcommand{\reals}{\mathbf{R}}
\newcommand{\nats}{\mbox{${\rm I\!N }$}}

\def\gat{\gamma_a^T}
\def\la{\lambda}
\def\om{\omega}
\def\La{\Lambda}
\def\lam{\lambda}
\def\al{\alpha}
\def\gam{\gamma}
\def\Th{\theta}
\def\chs{\chi^\star}
\def\dirac{D\!\!\!\!/}
\def\pip{\Pi_+}
\def\pim{\Pi_-}
\def\pipl{\Pi_+^\star}
\def\pimi{\Pi_-^\star}
\def\gf{\tilde{\gamma}}
\def\rand{\left|_{\partial M}\right. }
\def\ch{\cosh \Theta}
\def\sh{\sinh\Theta}
\def\nen{\Lambda \ch -2\om\sh}
\def\tr{\mbox{Tr}}
\def\ov{\over}
\def\id{1\!\mbox{l}}
\def\sign{{\rm sign}}
\def\Tr{{\rm Tr}}
\def\tr{{\rm tr}}

\newcommand{\ga}{\gamma_x}
\newcommand{\gt}{\gamma_5}
\newcommand{\gm}{\gamma_y}
\newcommand{\beq}{\begin{eqnarray}}
\newcommand{\eeq}{\end{eqnarray}}
\newcommand{\nn}{\nonumber}

\makeatletter
  \renewcommand{\theequation}{%
   \thesection.\arabic{equation}}
  \@addtoreset{equation}{section}
 \makeatother

\title{Spectral functions of the Dirac operator under
local boundary conditions}
\author{C.G. Beneventano\thanks{E-mail:
gabriela@obelix.fisica.unlp.edu.ar}\,,
E.M. Santangelo\thanks{E-mail: mariel@obelix.fisica.unlp.edu.ar}\\
Departamento de F\'{\i}sica, Universidad Nacional de La Plata\\
C.C.67, 1900 La Plata, Argentina} \maketitle
\begin{abstract}
After a brief discussion of elliptic boundary problems and their
properties, we concentrate on a particular example: the Euclidean
Dirac operator in two dimensions, with its domain determined by
local boundary conditions. We discuss the meromorphic structure of
the zeta function of the associated second order problem, as well
as the main characteristic of the first order problem, i.e., the
boundary contribution to the spectral asymmetry, as defined
through the eta function.
\end{abstract}
\section{Introduction}\label{intr}

Spectral functions are of interest both in quantum field theory
and in mathematics (for a recent review, see \cite{kirs01}). In
particular, $\zeta$-functions and heat kernels of elliptic
boundary problems are known to provide an elegant regularization
method \cite{zeta} for the evaluation of objects as one-loop
effective actions and Casimir energies, as discussed, for
instance, in the reviews \cite{eliz}.

In the case of operators with a non positive-definite principal
symbol, another spectral function has been studied, known as
$\eta$-function \cite{gilk95b}, which characterizes the spectral
asymmetry of the operator. This spectral function was originally
introduced in \cite{aps} (see also \cite{EGH}), where an index
theorem for manifolds with boundary was derived. In fact, the
$\eta$-function of the Dirac operator, suitably restricted to the
boundary, is proportional to the difference between the anomaly
and the index of the Dirac operator, acting on functions
satisfying nonlocal Atiyah-Patodi-Singer (APS) boundary
conditions. Some examples of application were discussed in
\cite{schroer,diskaps}.

Such nonlocal boundary conditions were introduced mainly for
mathematical reasons, although several applications of this type
of boundary value problems to physical systems have emerged,
ranging from one-loop quantum cosmology \cite{esp}, fermions
propagating in external magnetic fields \cite{bs} or so-called
$S-$branes, which are mapped into themselves under $T-$duality
\cite{vass}. So far, $\eta$-functions have found their most
interesting physical applications in the discussion of fermion
number fractionization \cite{nsem}: The fractional part of the
vacuum charge is proportional to $\eta(0)$. The $\eta-$function
also appears as a contribution to the phase of the fermionic
determinants and, thus, to effective actions \cite{cez}.
Furthermore, both the index and the $\eta$-invariant of the Dirac
operator are related to scattering data via a generalization of
the well-known Levinson theorem \cite{wirai}. A thorough
discussion of the index, $\zeta-$ and $\eta-$functions in terms of
boundary spectral functions for APS boundary problems can be found
in \cite{grub96-6-31,wojaps}.

Alternatively, one may consider the boundary value problem for the
Dirac operator acting on functions that satisfy local, bag-like,
boundary conditions. These conditions are closely related to those
appearing in the effective models of quark confinement known as
MIT bag models \cite{bag}, or their generalizations, the chiral
bag models \cite{chbag}. The physical motivation for studying
these local boundary conditions is thus clear.

Chiral bag boundary conditions
\cite{hras84-245-118,wipf95-443-201,durr} can be defined in any
even dimensional manifold. They contain a real parameter $\theta$,
which is to be interpreted as an analytic continuation of the well
known $\theta$-parameter in gauge theories. Indeed, for
$\theta\neq 0$, the effective actions for the Dirac fermions
contain a $CP$-breaking term proportional to $\theta$ and
proportional to the instanton number \cite{wipf95-443-201}.

For $\Th\neq0$ we will refer to the bag boundary conditions as
chiral while, in the particular case $\Th=0$, we will call them
non-chiral or pure MIT conditions. In both cases, the Euclidean
Dirac operator is self-adjoint. The ellipticity of the boundary
value problem is a more subtle point. In fact, even though the
boundary conditions are local for the first order problem, for
$\theta \neq 0$, the boundary conditions for the associated second
order problem are of mixed oblique type \cite{espo02-66-085014}
and, under certain circumstances, oblique boundary conditions are
not strongly elliptic \cite{dowk99-16-1917,avra99-200-495}.

In this work, we will concentrate on the Euclidean Dirac operator
in two dimensions, acting on functions satisfying local, chiral
bag, boundary conditions. The two-dimensional case, being the
simplest one, already contains all the main characteristics
appearing for the problem in higher dimensional cases
\cite{BGKS-03}.

We briefly review, in section \ref{Sect2}, the concepts of weak
and strong ellipticity of boundary value problems, and the
consequences of their validity. In section \ref{Sect3}, we prove
that the first order boundary problem at hand is indeed strongly
elliptic. In the same section, we define the associated second
order problem and prove that it is also strongly elliptic, even
though the boundary conditions involve tangential derivatives.

In section \ref{Sect4} we give an explicit expression for the heat
kernel of the second order problem in an infinite cylinder. By
making use of such expression we analyze, in section \ref{Sect5},
the meromorphic structure of the corresponding zeta function for
manifolds of product type.

One of the main characteristics of bag boundary conditions is that
they lead to an asymmetry in the non-zero spectrum. Note that, as
in any even dimension, there is no volume contribution to the
asymmetry (for a proof see, for instance, \cite{gilk95b};
qualitatively, this is due to the existence of $\gamma_5$, which
anticommutes with the Dirac operator). So, the boundary
contribution is also the total asymmetry. In section \ref{Sect6}
we study, again for product manifolds, the meromorphic structure
of the eta function and the boundary contribution to the spectral
asymmetry for chiral bag boundary conditions
\cite{bene02-35-9343}.

\section{Elliptic boundary problems}\label{Sect2}

In this section, we will briefly review the definition of weakly
(Lopatinski-Shapiro) and strongly elliptic boundary systems, and
state their main properties (for details, see \cite{gilk95b,
seel68-10-288, seel69-91-963}).

 Let $M$ a smooth compact manifold of dimension $n$, with a
smooth boundary $\partial M$.

In each local coordinate system, call $x=(x_1,...,x_{n-1})$ the
coordinates on $\partial M$. Let $y$ ($\in \mathbb{R}$) the
interior normal to the boundary. So, $z=(x,y)\in\mathbb{R}^n$.
Call $\mathbb{R}_+ ^n$ the half space $y\geq 0$. Consider, in
$\mathbb{R}_+ ^n$, the differential operator of order $m$, acting
on a $q$-dimensional complex vector bundle $V$:
\[P=\sum _{j=0}^m P_j (y)
D_{y}^{m-j}\,,\quad(D_y=-i\frac{\partial}{\partial y})\,,\] where
$P_j$ is a differential operator ($q\times q$ matrix) of order
$\leq j$ on $\mathbb{R}^{n-1}$.

Then, calling $(\xi,\tau )$ the symbolic variable corresponding to
$(x,y)$ we have, for the symbol of $P$:
\[\sigma (P)=\sum _j \sigma (P_j)(x,y,\xi ) {\tau}^{m-j}\,,\]
which is a polynomial of order $m$ in the $n$ variables
$(\xi,\tau)$.

The leading symbol is its $m$-th order part
\[{\sigma}_L(P)=\sum _j {\sigma}_j (P_j)(x,y,\xi ) {\tau}^{m-j}\,.\]

Moreover, we define a partial leading symbol, by:
\[{\sigma}_L^{\prime}(P)= \sum _j {\sigma}_j (P_j)(x,0,\xi )
{D_y}^{m-j}\,.\]

Suppose, near the boundary, we have certain given operators
(defining the boundary conditions)
\[B_j=\sum_{k=1}^m B_{jk} D_y ^{m-k}\,,\quad 1\leq j\leq \frac{m q}{2}\] where the $B_{jk}$
are a system of differential operators ($1\times q$ matrices)
acting on ${\mathbb{R}}^{n-1}$.

The collection of operators $P, B_1,..., B_{\frac{mq}{2}}$
constitute a weakly elliptic boundary system if $P$ is elliptic
and, for $g=(g_1,..., g_{\frac{mq}{2}})$ arbitrary, $x$ in
$\mathbb{R}^{n-1}$ and $\xi \neq 0$ in $\mathbb{R}^{n-1}$ there is
a unique solution to the following problem for $y>0$ : \beq \nn
{\sigma}_L^{\prime}(P)(x,\xi,D_y) u&=&0\\ \nn \lim _{y\rightarrow
\infty} u(y)&=&0 \\{\sigma}_L^{\prime} (B_j)(x,\xi,D_y) u&=&g_j
\quad {\rm at} \, \, y=0\quad {\rm
for}\,j=1,...,\frac{mq}{2}\,.\label{lop}\eeq This condition is
also known as the Lopatinski-Shapiro condition \cite{gilk95b}.
When this condition holds, an operator $P_B$ can be defined as the
operator $P$, acting on functions such that $B_ju\rfloor_{y=0}=0$.

\bigskip

Now, in order to meet some further requirements, one needs to
impose stronger conditions on the boundary system:

The collection $P, B_1,..., B_{\frac{mq}{2}}$ constitutes a
strongly elliptic boundary system in a cone ${\cal
K}\,\subset\,\mathbb{C}$ including the origin if

$i$) For $(\xi,\tau)\neq (0,0)$, ${\sigma}_L (P)$ has no
eigenvalue in ${\cal K}$ and

$ii$) For each $x$ and each $(\xi , \lambda )\neq (0,0)$, with
$\lambda\,\in\,{\cal K}$, the boundary problem \beq \nn
{\sigma}_L^{\prime}(P)(x,\xi,D_y) u&=&\lambda u\\ \nn \lim
_{y\rightarrow \infty} u(y)&=&0 \\{\sigma}_L^{\prime}
(B_j)(x,\xi,D_y) u&=&g_j \quad {\rm at} \, \, y=0,\quad
j=1,...,\frac{mq}{2}\label{sell}\eeq has a unique solution. (Note
this reduces to the Lopatinski-Shapiro condition for $\lambda=0$).
The cone ${\cal K}$ is known as Agmon's cone \cite{agmon}.

When the strong ellipticity condition holds, an approximation
(parametrix) to the resolvent $(P_B -\lambda )^{-1}$ can be found
\cite{seel69-91-963} and, from it, one can define the complex
powers
\[(P_B)^{-s}=\frac{i}{2\pi}\int_{\Gamma} {\lambda}^{-s}
(P_B-\lambda)^{-1} d\lambda\,,\] with $\Gamma$ an appropriate
curve in the cone where $(P_B-\lambda)^{-1}$ is known to exist.

The coefficients in the expansion of the parametrix (Seeley's
coefficients) must satisfy the condition of representing an
inverse for $P_B-\lambda$, and they must also adjust the boundary
condition. As a consequence, both bulk and boundary Seeley's
coefficients appear.

If, moreover, the leading symbol of the operator is
positive-definite, another operator can be defined as

\beq K(t,P_B)=e^{-tP_B}=\frac{i}{2\pi}\int_{\Gamma} e^{-\lambda t}
(P_B-\lambda)^{-1} d\lambda \quad,\,t>0 \,.\eeq

It can be shown that its kernel $K(z,z';t)$ is the fundamental
solution of the heat equation, and is known as the heat kernel of
$P_B$.

We will be interested in the traces of the complex powers and of
the heat kernel.

In addition, let $P_B$ self-adjoint. Then, the following theorem
is known to hold (see Theorem 1.11.3 in \cite{gilk95b}):

a) There exists a discrete spectral resolution $\{\phi_{\nu},
\lambda_{\nu}\}$ of $P_B$.

b) There exist $\nu_0$ and $\delta>0$, so that if $\nu\geq\nu_0$,
then $|\lambda_{\nu}|\geq {\nu}^{\delta}$.

c) If the boundary value problem defining $P_B$ is strongly
elliptic, only a finite number of the $\lambda_{\nu}$ are
negative.

Whenever the principal symbol of the operator is
positive-definite, the operator $K(t,P_B)$ is infinitely
smoothing, and its trace, in the basis of eigenfunctions is given
by

\beq Tr K(t,P_B)=\sum_{\nu} e^{-{\lambda}_{\nu} t}\,.\eeq

Moreover, if $P_B$ is positive-definite (i.e., there are no
negative or zero eigenvalues in the spectral resolution), one
further defines its $\zeta$ function as $\zeta(s,P_B)=Tr
(P_B^{-s})$,

\beq \zeta(s,P_B)=\sum_{\nu} ({\lambda}_{\nu})^{-s}\,.\eeq

Both spectral functions are related through a Mellin transform,

\beq\nn \zeta(s,P_B)&=&\frac{1}{\Gamma(s)} \int_0^{\infty}
dt\,t^{s-1} \sum_{\nu}
e^{-{\lambda}_{\nu}t}\\&=&\frac{1}{\Gamma(s)} \int_0^{\infty}
dt\,t^{s-1}TrK(t,P_B)\,.\eeq

\bigskip

\section{Chiral bag boundary conditions}\label{Sect3}

\subsection{First order boundary problem}

In this section, we will discuss, in particular, the properties of
the boundary value problem defined by the Euclidean Dirac operator
on a two-dimensional manifold, acting on spinors satisfying local
(chiral bag) boundary conditions \cite{wipf95-443-201}.

To this end, it is convenient to choose a chiral representation
for the Euclidean $\gamma$-matrices in two dimensions, \beq
\ga=\sigma_1,\qquad\gm=\sigma_2\qquad {\rm and}\qquad
\gt=-i\ga\gm=\sigma_3.\label{gammatr}\eeq

Then, the free Dirac operator acts on two-component spinors
$\psi$, and it takes the form \beq
P=i(\ga\partial_x+\gm\partial_y)=\left(\matrix{
  0&\partial_y+A \cr
  -\partial_y + A & 0 } \right),\label{op}\eeq
where $A$ is the operator $A=i\partial_x$, which will play an
important role in what follows.

 The euclidean ``time"-coordinate $x$ is tangential
to the boundary at $y=0$. The ``spatial" variable $y\geq 0$ is
normal to the boundary and grows toward the interior of the
manifold. The projector defining the local chiral bag boundary
condition \beq
\left.\pim\psi\right\rfloor_{y=0}=0\label{bcdir}\eeq at the
boundary $y=0$ reads \beq \pim=\frac12(\textbf{1}-i\gt \gm e^{-\gt
\Th})=\frac12(\textbf{1}-\ga e^{-\gt \Th})
=\frac12\left(\begin{array}{cc}
  1 & -e^{\Th} \\
  -e^{-\Th} &1 \\
\end{array}\right)\,.\label{proj1}\eeq

In the case at hand, with the notation of section \ref{Sect2},
$n=2, m=1, q=2$.

The leading symbol of (\ref{op}) is given by \beq
\sigma_L(P)=\left(\matrix{
  0&-\xi+i\tau\cr
  -\xi-i\tau & 0 } \right)\,,\label{ls}\eeq
which is easily seen to be invertible for $(\xi,\tau)\neq(0,0)$.

Now, in order to analyze the ellipticity of the boundary value
problem, we evaluate that \beq\sigma_L^{\prime}(P)=\left(\matrix{
  0&-\xi+\partial_y\cr
  -\xi-\partial_y & 0 } \right)\,,\label{pls}\eeq
and the boundary operator ($1\times 2$ matrix)\beq
B_1=B_{11}\,{\partial_y}^0=\frac12(\begin{array}{cc}
  1 & -e^{\Th} \\
\end{array})\,.\label{b1}\eeq

Note that, this being a multiplicative operator, its partial
leading symbol coincides with the operator itself.

We first prove that the boundary value problem is weakly elliptic.
The first two equations in (\ref{lop}) give \beq
u(\xi,y)=\left(\begin{array}{c}
  u_1(\xi,y) \\
  u_2(\xi,y) \\
\end{array}\right)=\left(\begin{array}{c}
  C_1(\xi)\Theta(\xi) e^{-\xi y}\\
  C_2(\xi) \Theta(-\xi) e^{\xi y}\\
\end{array}\right)\,,\label{spinor}\eeq
where $\Theta(\xi)$ is the Heaviside function.

Now, the third equation in (\ref{lop}) univocally determines $
C_1(\xi)=2g(\xi)$, for $\xi>0$, and $ C_2(\xi)=-2g(\xi)e^{-\Th}$,
for $\xi<0$. This shows that weak ellipticity does hold.

As for strong ellipticity, condition $i)$ in section \ref{Sect2},
together with (\ref{ls}), lead to
\[\lambda^2-(\xi^2+\tau^2)\neq 0\,,\]
which is satisfied if $\lambda\in {\cal K}=\mathbb{C}-\mathbb{R}^+
-\mathbb{R}^-$.

Moreover, the first two conditions in (\ref{sell}) lead to \beq
u(\xi,y)=\left(\begin{array}{c}
  u_1(\xi,y) \\
  u_2(\xi,y) \\
\end{array}\right)=\left(\begin{array}{c}
  C_1(\xi) e^{-\mu y}\\
  C_1(\xi)\frac{(\mu-\xi)}{\lambda} e^{-\mu y}\\
\end{array}\right)\,,\label{spinor2}\eeq
where $\mu^2=\xi^2-\lambda^2$, and the condition $\lambda\in{\cal
K}$ allows to choose $\Re(\mu)>0$.

Now, the equation for the symbol of the boundary operator in
(\ref{sell}) gives
\[C_1(\xi)[\lambda-e^{\Th}(\mu-\xi)]=2g(\xi)\lambda\,.\]

So, the problem has a unique solution if $\lambda\neq
-\frac{\xi}{\cosh{\Th}}$, a condition which is satisfied for
$\lambda\in{\cal K}$. This proves the strong ellipticity of the
first order boundary problem with respect to
$\mathbb{C}-\mathbb{R}^+ -\mathbb{R}^-$.

\subsection{Associated second order problem}

To the first order boundary problem defined by equations
(\ref{op}) and (\ref{bcdir}), a second order problem is naturally
associated. This last consists of the Laplacian \beq
P^2=\left(\matrix{
  -\partial^2_y+A^2 &0\cr
  0&-\partial^2_y + A^2} \right),\label{op2}\eeq
acting on a two-dimensional vector bundle, whose elements satisfy
the boundary
conditions \beq\nn\left.\pim\psi\right\rfloor_{y=0}&=&0\\
\left.\pim P\psi\right\rfloor_{y=0}&=&0\,.\label{bclapl}\eeq

In equation (\ref{op2}), $A$ is defined as for the first order
problem.

Although the ellipticity of the second order problem follows from
the ellipticity of the corresponding first order one and Lemma
1.11.2 (b) in \cite{gilk95b}, we prefer to give a second proof
showing that the boundary operator involves tangential
derivatives. This proof makes apparent that the chiral boundary
conditions are non-standard (oblique) boundary conditions.

With the notation of section \ref{Sect2}, $n=2$, $q=2$ and $m=2$.
So, we have a set of two boundary operators ($1\times 2$
matrices), given by \beq\nn B_1&=&B_{11} {\partial_y}^0=\frac12
\left(\begin{array}{cc}
  1 & -e^{\Th}\\
\end{array}\right)\\B_2&=&B_{21} {\partial_y}^1+B_{22}
{\partial_y}^0=\frac12\left[\left(\begin{array}{cc}
  e^{\Th} & 1 \\
\end{array}\right){\partial_y}+\left(\begin{array}{cc}
  -e^{\Th} & 1 \\
\end{array}\right)i\partial_x\right]\,.\eeq

As is well known, the Laplacian ($P^2$) is an elliptic operator.
Its leading symbol is given by \beq \sigma_L (P^2)=\left(\matrix{
  \tau^2+\xi^2 &0\cr
  0&\tau^2+\xi^2} \right)\,.\label{lsp2}\eeq

  It's easy to check that it has no eigenvalue for
  $(\xi,\tau)\neq(0,0)$, and \hfill\break $\lambda\in {\cal K}=\mathbb{C}-\mathbb{R}^+$.

  Now, the partial leading symbols are
\beq \sigma_L^{\prime} (P^2)=\left(\matrix{
  \xi^2-{\partial_y}^2 &0\cr
  0&\xi^2-{\partial_y}^2} \right)\,,\label{plsp2}\eeq
  and
  \beq\nn \sigma_L^{\prime} (B_1)&=&B_1
\\\sigma_L^{\prime}(B_2)&=&\frac12\left[\left(\begin{array}{cc}
  e^{\Th} & 1 \\
\end{array}\right){\partial_y}+\left(\begin{array}{cc}
  -e^{\Th} & 1 \\
\end{array}\right)(-\xi)\right]\,.\label{plsb}\eeq

For this second order boundary problem, weak ellipticity is but a
particular case of the strong one. So, let us prove this last
holds.

The first two equations in (\ref{sell}) lead to \beq
u(\xi,y)=\left(\begin{array}{c}
  u_1(\xi,y) \\
  u_2(\xi,y) \\
\end{array}\right)=\left(\begin{array}{c}
  C_1(\xi) e^{-\mu y}\\
  C_2(\xi) e^{-\mu y}\\
\end{array}\right)\,,\label{spinor3}\eeq
where $\mu^2=\xi^2-\lambda$, and the condition $\lambda\in{\cal
K}$ allows to choose $\Re(\mu)>0$.

When the last equation in (\ref{sell}) is used, together with
(\ref{plsb}), one can see that the problem has a unique solution
whenever $\xi^2\neq \lambda \cosh^2{\Th}$, a condition which is
satisfied if $\lambda\in {\cal K}$, which proves strong
ellipticity with respect to $\mathbb{C}-\mathbb{R}^+$. Moreover,
if $\lambda=0$, the condition is fulfilled for $\xi\neq 0$, which
proves weak ellipticity.

\section{The heat kernel in an infinite cylinder}
\label{Sect4}

In this section, we will obtain the explicit expression for the
heat kernel of the second order problem in an infinite cylinder.
This will be the main ingredient in the subsequent study of
spectral functions on a compact manifold.

Let us start by discussing some properties of the boundary
projector \beq \pim=\frac12(\textbf{1}-i\gt \gm e^{-\gt \Th})\eeq
and of its orthogonal projector \beq \pip=\frac12(\textbf{1}+i\gt
\gm e^{-\gt \Th})\,.\eeq

Note that these projectors are not self-adjoint (except for the
particular case $\Th =0$). Rather, one has
\beq\nn\pimi&=&\frac12(\textbf{1}-i\gt \gm e^{\gt \Th})\\
\pipl&=&\frac12(\textbf{1}+i\gt \gm e^{\gt \Th})
\,,\label{pis}\eeq  and the following equations hold

\beq\begin{array}{l} \pipl \pip=\cosh{(\Th \gt)}\exp{(-\Th \gt)}
\pip=\pipl \cosh{(\Th \gt)}\exp{(-\Th \gt)}\nn\\\pimi
\pip=\sinh{(\Th \gt)}\exp{(-\Th \gt)} \pip=\pimi \sinh{(\Th
\gt)}\exp{(-\Th \gt)}\nn\\\pimi \pim=\cosh{(\Th \gt)}\exp{(-\Th
\gt)} \pim=\pimi \cosh{(\Th \gt)}\exp{(-\Th \gt)} \nn\\\pipl
\pim=\sinh{(\Th \gt)}\exp{(-\Th \gt)} \pim=\pipl \sinh{(\Th
\gt)}\exp{(-\Th \gt)}\,.\end{array}\label{prod}\eeq

\bigskip

 We use $\Pi _-$ to define boundary conditions for $P$, as in equation (\ref{bcdir}).
Similarly, we shall let the equations (\ref{bclapl}) define the
associated boundary condition for $P^2$. We wil call this second
order boundary value problem $(P^2,{\cal B})$.

In what follows, we present the heat kernel for $(P^2,{\cal B})$
in an infinite cylinder ${\cal M}= \mathbb{R} _+ \times {\cal N}$,
where ${\cal N}$ is the closed boundary.

In order to determine the heat kernel, it is useful to note that
the chiral bag boundary conditions in equations (\ref{bclapl})
are equivalent, for each eigenvalue of the tangential part $A$ of
the operator $P$, to Dirichlet boundary conditions on part of the
fibre, and Robin (modified Neumann) on the rest.

In fact, let's first notice that the operators ${\cal
P}_+=\frac{\pip \pipl}{\cosh^2{\Th}}$ and ${\cal P}_-=\frac{\pimi
\pim}{\cosh^2{\Th}}$ are self-adjoint projectors, and they satisfy
${\cal P}_++{\cal P}_-=\textbf{1}$, splitting $V$ into two
complementary subspaces.

As before, let $x$ be the coordinate on the boundary and $z=(x,y)$
a local coordinate system on the manifold. If we call
$\phi_{\omega}(x)=e^{i\omega x}$ the eigenfunctions of the
operator $A=i\partial_x$ corresponding to the eigenvalue $\omega$,
we can expand $\psi(x,y)=\sum_{\omega}f_{\omega}(y)
\phi_{\omega}(x)$. If $\psi={\cal P}_+ \psi$, then the first
condition in equation (\ref{bclapl}) is identically satisfied, and
only the second condition in (\ref{bclapl}) must be imposed at the
boundary which, for each $\omega$, reduces to
\[\cosh{\Th}e^{-\Th \gt}\left(\partial_y +\omega \tanh{\Th}\right)f_{\omega}\rfloor_{y=0}=0\,.\]
Since the factor to the left of the parenthesis is invertible,
this is nothing but a Robin boundary condition.

In the subspace $\psi={\cal P}_- \psi$, the boundary equations
reduce, respectively, to\[\cosh{\Th}e^{\Th
\gt}f_{\omega}\rfloor_{y=0}=0\,,\] and\[\omega
f_{\omega}\rfloor_{y=0}=0\,.\] Thus, in this subspace, the
requirement that both conditions are simultaneously satisfied
leads to homogeneous Dirichlet boundary conditions.

As a consequence, the complete heat kernel can be written as a
Dirichlet heat kernel on ${\cal P}_-V$ and a Robin heat kernel on
${\cal P}_+ V$. For the convenience of the reader we make the
single ingredients explicit \cite{cars86b} and write, with
$\rho=y-y'$ and $\eta=y+y'$, \beq K(z,z';t) &=& K (z,z';t) ( {\cal
P} _- +
{\cal P} _+) \nn\\
&=& \frac 1 {\beta\sqrt{4\pi
t}}\sum_{\omega}\phi_{\omega}^{\star}(x') \phi_{\omega}(x)
e^{-\omega^2
t}\left(e^{\frac{-\rho^2}{4t}}-e^{\frac{-\eta^2}{4t}}\right){\cal
P}_- \nn\\
&+&\frac 1 {\beta\sqrt{4\pi
t}}\sum_{\omega}\phi_{\omega}^{\star}(x') \phi_{\omega}(x)
e^{-\omega^2
t}\left\{\left(e^{\frac{-\rho^2}{4t}}+e^{\frac{-\eta^2}{4t}}\right)\right.
\nn\\
& &\left.+2 \sqrt{\pi t}\,\, \omega \tanh \theta e^{\omega ^2 t
\tanh ^2 \theta - \omega \eta\tanh \theta } erfc [ u_\omega (\eta
, t)] \right\} {\cal P}_+ \nn\\ &=&\frac{1}{\beta\sqrt{4\pi
t}}\sum_{\omega}e^{i\omega (x-x')} e^{-\omega^2
t}\left\{\left(e^{\frac{-\rho^2}{4t}}-e^{\frac{-\eta^2}{4t}}\right)
\mathbf{1}\right.\label{hk}\\
& & \left.+\frac{2\pip \pipl}{\cosh^2(\Th)}\left[1+\sqrt{(\pi
t)}\omega \tanh{\Th}e^{u_{\omega}(\eta,t)^2}
erfc[u_{\omega}(\eta,t)]\right]e^{\frac{-\eta^2}{4t}}\right\}
\nn\eeq where $\beta=\mbox{Vol} ({\cal N})$,
$u_{\omega}(\eta,t)=\frac{\eta}{\sqrt{4t}}-\sqrt{t}\omega\tanh(\Th)$,
and\[erfc(x)=\frac{2}{\sqrt{\pi}}\int_{x}^{\infty}d\xi
e^{-\xi^2}\,\] is the complementary error function.

The heat kernel in equation (\ref{hk}) was first presented in
\cite{bene02-35-9343}. For the particular case of an antiperiodic
boundary fiber, it coincides with the Fourier transform of
equation (101) in \cite{wipf2}.

In the appendix \ref{a1} we show that the asymptotics of the heat
kernel in a compact manifold of product type coincides with the
asymptotics of (\ref{hk}).

\section{Meromorphic properties of the zeta function}
\label{Sect5}

Let us now analyze the boundary contributions to the global zeta
function related to (\ref{hk}). We first note that global
quantities are necessarily divergent due to the non-compact nature
of our manifold ${\cal M}$. It is easy to see that such
divergencies come, upon integration over $\mathbb{R}^+$, from the
first term in (\ref{hk}). As shown in appendix \ref{a1}, this term
(duely smeared) completes the volume contribution in a compact
manifold.

In the following we will, without changing the notation, ignore
this term, and this will allow us to determine the boundary
contributions to the global zeta function.

Let us then consider the trace of ({\ref{hk}) ignoring the first
term. It is convenient to perform the Dirac trace ($tr$) first.
Since
\[tr \left(\frac{2\pip \pipl}{\cosh^2(\Th)}\right)=2\,,\] the trace
of the 'boundary' heat kernel reduces to \beq Tr K&=&
\sum_{\omega} \omega
\tanh{\Th}e^{-\omega^2 t}\nn\\
& &\times \,\,\int_0^{\infty} dy\, erfc[u_{\omega}(2y ,
t)]e^{\frac{-y^2}{t}+ u_\omega ^2 (2y ,\,\, t) }\,,\nn\eeq where
the second and third term in (\ref{hk}) have cancelled each other.
Now, using that \[ -\frac12 \frac{\partial}{\partial y}
\left[e^{-y^2/t+u_{\omega}^2(2y,t)}
erfc\left[u_{\omega}(2y,t)\right]\right] =\]\beq
e^{-y^2/t}\left[\frac{1}{\sqrt{\pi t}}+\omega \tanh\Th\,
e^{u_{\omega}^2(2y,t)}
erfc\left[u_{\omega}(2y,t)\right]\right]\,,\label{ident}\eeq we
get
\[ Tr K= \frac12\sum_{\omega}e^{-\omega^2
t}\left[e^{u_{\omega}^2(0,t)}erfc\left[u_{\omega}(0,t)\right]-1\right]=\]\beq
\frac12 \sum_{\omega}\left[e^{\frac{-\omega^2 t
}{\cosh^2{\Th}}}\left[1+erf(\omega\sqrt{t}
\tanh{\Th})\right]-e^{-\omega^2 t}\right]\,.\nn\eeq Here we used
$erf(x)=-erf(-x)=1-erfc(x)$.

Now, we can Mellin transform this trace, to obtain the 'boundary'
zeta function of the square of the Dirac operator in the infinite
cylinder \beq
\zeta(s,P^2)&=&\frac{1}{2\Gamma(s)}\sum_{\omega}\int_0^{\infty}dt
\,\, t^{s-1}\left[e^{\frac{-\omega^2
t }{\cosh^2{\Th}}}-e^{-\omega^2 t}\right]\nn\\
& &+
\frac{1}{2\Gamma(s)}\sum_{\omega}\int_0^{\infty}dt\,\,t^{s-1}e^{\frac{-\omega^2
t }{\cosh^2{\Th}}}erf(\omega\sqrt{t}
\tanh{\Th})\nn\\
&=&\zeta_1(s,P^2)+\zeta_2(s,P^2)\,.\label{z}\eeq

The first contribution can be readily seen to be \beq
\zeta_1(s,P^2)=\frac{1}{2}\left(\cosh^{2s}
{\Th}-1\right)\zeta(s,A^2)\,.\label{z1}\eeq

As for the second contribution to (\ref{z}), it is given by \beq
\zeta_2(s,P^2)=\frac{1}{2\Gamma(s)}\sum_{\omega}\int_0^{\infty}dt\,\,t^{s-1}e^{\frac{-\omega^2
t }{\cosh^2{\Th}}}\frac{2}{\sqrt{\pi}}\int_0^{(\omega\sqrt{t}
\tanh{\Th})} d\xi e^{-\xi^2}\,.\nn\eeq

After changing variables according to
$y=\frac{\xi\cosh\Th}{\sqrt{t}\omega}$, and interchanging
integrals, one finally gets \beq
\zeta_2(s,P^2)&=&\frac{\Gamma\left(s+\frac{1}{2}\right)}{2\Gamma(s)}\cosh^{2s}
\Th
\sum_{\omega}sign(\omega) \left(\omega^2\right)^{-s}\nn\\
& &\quad \quad \quad \quad \quad \quad \times \,\,
\frac{2}{\sqrt{\pi}}\int_0^{\sinh{\Th}} dy
\left(1+y^2\right)^{-s-\frac{1}{2}}\nn\\
&=& \frac{\Gamma\left(s+\frac{1}{2}\right)}{2\Gamma(s)}\cosh^{2s}
\Th\eta(2s,A) \frac{2}{\sqrt{\pi}}\int_0^{\sinh{\Th}}
dy\left(1+y^2\right)^{-s-\frac{1}{2}}\nn\\
&=&\frac 1 { \sqrt{\pi}} \frac{\Gamma\left( s + \frac  1 2 \right)
} {\Gamma (s)} \sinh \theta \cosh ^{2s} \theta \eta (2s,
A)\nn\\
& &\quad \times \,\,_2F_1 \left( \frac 1 2 , \frac 1 2 +s, \frac 3
2 ; -\sinh ^2 \theta \right)\,. \label{z2}\eeq The structure of
the zeta function is similar to the structure found for spectral
boundary conditions, see e.g. \cite{grub96-6-31}. In particular,
the analysis of the zeta function on ${\cal M}$ has been reduced
to the analysis of the zeta and eta functions on the boundary
${\cal N}$.

As already commented, from (\ref{z1}) and (\ref{z2}) one can
determine the positions of the poles and corresponding residues
for the zeta function in any cylindrical product manifold, in
terms of the meromorphic structure of the zeta and eta functions
of the operator $A$. For the rightmost poles, explicit results can
be given in terms of the geometry of the boundary. For example,
for $s=1/2$ we see that \beq \mbox{Res } \zeta_1 \left( \frac {1}
2 , P^2 \right) &=& \frac 1 2 \left( \cosh{\theta} -1 \right)
\mbox{Res } \zeta \left(
\frac {1} 2 , A^2 \right) \nn\\
&=& \frac {\beta} 2 \left( \cosh {\theta} -1 \right) \frac{ (4\pi
) ^{-1/2} } {\Gamma \left( \frac{1} 2 \right)}.\nn\eeq Because
$\zeta_2$ does not contribute, given $\eta (2s,A)$ is regular at
$s=1/2$ \cite{gilk95b}, this equals $\mbox{Res }\zeta (s,P^2)$ and
is the result expected from the calculation on the ball
\cite{kirs02-104-119}. For $\theta =0$ the residue disappears, as
is known to happen for non-chiral local bag boundary conditions
\cite{gilk95b}. Further results can be obtained by using Theorem
4.4.1 of \cite{gilk95b}. For the particular case of $s=0$, the
fact that $\zeta (s,A^2)$ and $\eta (2s,A)$ are regular at $s=0$
shows that $\zeta (0,P^2) =\zeta(0,P)=0$.

\section{Eta function and boundary contribution to the spectral asymmetry}
\label{Sect6}

As already commented, since the euclidean space-time we are
considering is even dimensional, there is no bulk contribution to
the asymmetry. To obtain the boundary contribution, the eigenvalue
problem for the Dirac operator $P$ should be investigated on a
collar neighborhood of the boundary. Here, we consider instead the
operator on the semi-infinite cylinder extending to $y\rightarrow
\infty$. As is well-known \cite{woj3}, since we are treating a
self-adjoint problem, this yields the correct answer for an
invertible boundary operator $A$. We shall discuss the non
invertible case toward the end of this section. Hence, for the
moment, we assume that $A$ has no zero modes.

Denoting the (real) eigenvalues of the Dirac operator by $\lam$,
the relevant spectral function is the eta function\beq \nn
\eta(s,P)&=&\sum_{\lam}\frac{\sign\lam} {|\lam|^s}
=\zeta\left(\frac{s+1}{2},P^2,P\right)\\&=& {1\ov
\Gamma\left(\frac{s+1}{2}\right)}
\int_0^{\infty}dt\,t^{\frac{s-1}{2}} \,\Tr
\left(P\,e^{-tP^2}\right)\,.\label{eta1}\eeq

In our particular case, using (\ref{op}) and (\ref{hk}), taking
Dirac traces, going to the diagonal, and integrating over the
tangential variable, one is left with \cite{bene02-35-9343}

\beq\nn &&\Tr \left(Pe^{-tP^2}\right)=\sum_{\omega}\omega
e^{-\omega ^2 t}\times\\& &\int\limits_0^{\infty}dy \left\{{1\ov
\sqrt{\pi t}}+ \omega\tanh\Th\, e^{u_n^2(2y,t)}\,
erfc\left[u_n(2y,t)\right]\right\}\,
{e^{-y^2/t}\ov\cosh\Th}\,.\label{t1}\eeq

Now, we may use the simple identity in equation (\ref{ident}) to
rewrite the relevant trace as follows,

\beq \nn\Tr \left(Pe^{-tP^2}\right)&=&- \sum_{\omega}\frac{\omega
e^{-\omega ^2 t/\cosh^2\Th}}{2\cosh\Th}\int_0^{\infty}dy
{\partial\ov
\partial y}\left[e^{-2y\omega \tanh\Th}
erfc\left(u(2y,t)\right)\right]
\\&=&\frac12 \sum_{\omega} {\omega\ov \cosh\Th} e^{-\omega ^2
t/\cosh^2\Th} erfc\left[-\sqrt{t}\tanh{\Th}\omega
\right]\,.\label{trace}\eeq

The boundary contribution to the eta function is obtained by
inserting (\ref{trace}) into (\ref{eta1}) and, hence, it is given
by

\beq \nn \eta(s,P) &=&\frac{1}{\Gamma(\frac{s+1}{2})}\sum_{\omega}
{\omega \ov
2\cosh{\Th}}\times\\&&\int_0^{\infty}dt\,t^{\frac{s-1}{2}}
e^{-\frac{\omega^2 t}{\cosh^2\Th}} \left[1-erf(-\sqrt{t}
\tanh{\Th}\omega )\right]\,.\eeq

Finally, changing variables to $\tau=\omega ^2 t/\cosh^2{\Th}$,
interchanging the order of the integrations and integrating over
$\tau$ one obtains the following rather explicit expression \beq
\nn\eta(s,P)&=&\frac12 \cosh^s\theta\sum_{\omega}
\left(\omega^2\right)^{-s/2} \left[\sign(\omega)+I(s,\theta)
\right]\\ &=& \frac12\cosh^s\theta \left[\eta(s,A)+
\zeta(\frac{s}{2},A^2)I(s,\theta)\right]\,,\label{etan}\eeq where
we have introduced the function \beq I(s,\theta)={2\sinh{\Th}\ov
\sqrt{\pi}}{\Gamma(\frac{s}{2}+1)\ov \Gamma(\frac{s}{2}+\frac12 )}
\,\,_2F_1 \left( \frac 1 2 , 1 +\frac{s}{2}, \frac 3 2 ; -\sinh ^2
\theta \right)\,.\eeq

Again, the analysis of the eta function has been reduced to the
analysis of the zeta and eta functions on the boundary
$\mathcal{N}$.

 To determine the spectral asymmetry, we particularize to $s=0$. With $\pi I(0,\Th)=2\arctan(\sinh\Th)$ we obtain, for $s=0$,
\beq \eta(0,P)=\frac12 \left\{\eta(0,A)+{2\ov\pi}\,
\zeta(0,A^2)\arctan(\sinh\Th)\right\}\,. \label{asyme}\eeq

Now, the second term within the curly brackets can be seen to
vanish, since the boundary is a closed manifold of odd
dimensionality. In fact, in our case, $\zeta(0,A^2)=a_1(A^2)=0$,
where $a_1(A^2)$ is a heat kernel coefficient in the notation of
\cite{gilk95b} (for details, see Theorem 1.12.2 and Lemma 1.8.2 in
this reference), and we are left with \beq \eta(0,P)=\frac12
\eta(0,A)\,.\label{asym}\eeq

As far as $A$ is invertible, this is the main result of this
section, relating the $\eta-$invariant of the Dirac operator to
the same invariant of the boundary operator. Note that the outcome
is the same irrespective of the value of $\theta$, i.e., it holds
both for pure MIT and chiral bag conditions. The first case was
treated before in \cite{woj3}.

As already pointed, (\ref{asym}) gives the whole spectral
asymmetry when the operator $A$ is invertible. In fact, for such
cases it was proved in \cite{woj3} (see also \cite{woj1}) that the
asymmetry splits, in the adiabatic (infinite volume) limit, into
the bulk contribution plus the infinite cylinder one. Moreover,
reference \cite{muller} shows that the spectral asymmetry is
independent from the size of the manifold when the boundary value
problem is self adjoint, as in our case. This, together with the
vanishing of the volume contribution in even dimensions, leads to
the previous conclusion.

\bigskip

Now, we study the more subtle case of a non-invertible boundary
operator $A$. Then, as can be seen from (\ref{t1}), $\omega =0$
would give no extra contribution in the semi-infinite cylinder.
However, in this case, the trace (\ref{trace}) can differ in a
substantial way from the corresponding one in the collar
neighborhood. As explained in \cite{woj1}, both large $t$
behaviors may be different, thus giving extra contributions to the
asymmetry  in the collar. This difference in high $t$ behavior is
due to the presence of ``small" eigenvalues, vanishing as the
inverse of the size of the manifold in the adiabatic limit
\cite{park}. These extra contribution can be determined, modulo
integers, by using the arguments in \cite{gilk95b,muller,woj2}. To
this end, consider the one-parameter family of differential
operators \beq P_{\al}=P+\frac{2\pi}{\beta}\al\gamma_{x},\qquad
P_0=P\,.\eeq

These operators share the same $\al$-independent domain. They are
invertible for $\al \neq 0$ and can be made invertible for all
$\al$ by subtracting the projector on the subspace of small
eigenvalues related to the zero-modes at $\al=0$. This then yields
a new family of operators $P'_{\al}$ and one obtains \beq
\eta(0,P_{\al})=\eta(0,P^{\prime}_{\al})\,\, {\rm
mod}\,\mathbb{Z}\qquad {\rm and} \qquad {d\ov
d\al}\eta(0,P_\al)={d\ov d\al}\eta(0,P'_\al)\,.\eeq Then,
differentiating with respect to $\al$ one finds for the spectral
flow \beq\nn {d\ov d\al}\eta(0,P'_\al)&=& \left.{1\ov
\Gamma(\frac{s+1}{2})}{d\ov d\al}
\int_0^\infty\,dt\,t^{\frac{s-1}{2}} \Tr\left(P'_\al
e^{-t{P'_\al}^2}\right)\right\rfloor_{s=0}\\ \nn &=& \left.{1\ov
\Gamma(\frac{s+1}{2})} \int_0^\infty\,dt\,t^{\frac{s-1}{2}}
\Tr\left[{d P'_\al\ov d\al}\left(1+ 2t{d\ov dt}
\right)e^{-t{P'_\al}^2}\right]\right\rfloor_{s=0}\\ \nn &=& \left.
-{2\pi\, s\ov
\beta\,\Gamma(\frac{s+1}{2})}\int_0^\infty\,dt\,t^{\frac{s-1}{2}}
\Tr\left(\gam_{x} e^{-t{P'_\al}^2}\right) \right\rfloor_{s=0}\\
&&\left.+{4\pi\ov \beta\Gamma(\frac{s+1}{2})}
\,\Tr\left(t^{\frac{s+1}{2}}\gam_{x}
e^{-t{P'_\al}^2}\right)_{t=0}^\infty
\right\rfloor_{s=0}\,,\label{sflux}\eeq where we performed a
partial integration to arrive at the last equation. In addition,
we used $dP'_\al/d\al=\frac{2\pi}{\beta}\gam_{x}$. Since
$P'_{\al}-P_\al$ is an operator of finite range we may safely skip
the prime in the last line of the above formula. Finally, the very
last term in equation (\ref{sflux}) can be seen to vanish, which
gives, for the spectral flow
\beq\nn {d\ov
d\al}\eta(0,P^{\prime}_\al)&=&-{\pi\ov\beta}{\rm Res}\rfloor_{s=0}
\left[\zeta(\frac{s+1}{2},A^2)\right.\\&&\left.+ {2\ov\pi}\eta(s+
1,A) \arctan(\sinh\Th)\right]\,.\label{sflux2}\eeq

Now, the second term can be seen to vanish, since (again with the
notation of \cite{gilk95b}), $\sqrt{\pi}\,{\rm
Res}\rfloor_{s=0}\,\eta(s+ 1,A)=2a_{0}(A^2,A)=0$.

Moreover, $\sqrt{\pi}\,{\rm Res}|_{s=0}\,\zeta(\frac{s+
1}{2},A^2)=2a_{0}(A^2)={\beta\ov \pi}$. Thus, one finally has for
the spectral flow, no matter whether $A$ is invertible or not \beq
{d\ov d\al}\eta(0,P_\al)=-1\,.\label{sflux3}\eeq

\bigskip

\bigskip

To summarize, this work shows that, in two dimensions, the first
order boundary value problem defined by chiral bag boundary
conditions is strongly elliptic, as is its associated second order
problem. We have related the meromorphic structure of the spectral
functions and the spectral asymmetry of the problem with the
meromorphic structure and spectral asymmetry of the operator $A$
(acting on the variable along the boundary), much in the way it
was done in \cite{grub96-6-31} for APS boundary conditions. Note
that, even though throughout this paper we have studied the free
Dirac operator, everything we've done holds for a gauge field such
that $A_x=A_x(x)$ (independent of the normal variable) and $A_y=0$
for, then, a gauge transformation can be performed of the form
$\psi^{\prime}(x,y)=e^{i\int_0^{x}A_x(x')dx'}\psi(x,y)$, which
leads to just a twist in the boundary fiber.

In this work, we have studied the case of a two-dimensional
manifold. However, similar results concerning the elliptic
character of the problem and the meromorphic properties of the
zeta function have been shown to hold for higher even dimensions
in \cite{BGKS-03}.

$\phantom{aa}$\\

\begin{appendix}

\section{Appendix}
\label{a1}

Since $(P^2,{\cal B})$ constitutes an elliptic, self adjoint,
boundary problem, with positive definite principal symbol, its
heat kernel can be defined; in the subspace orthogonal to zero and
negative modes (which, in principle, might exist in finite number)
it is an infinitely smoothing operator, and admits an asymptotic
expansion for $t\rightarrow 0$. Moreover, in our case, the
associated first order problem is also self adjoint. So,
$P^2_{\cal B}$ is non negative. As for zero modes, they have been
shown to be absent (at least for a simply connected boundary) in
reference \cite{wipf95-443-201}. Thus, the trace of the heat
kernel decreases exponentially for $t\rightarrow \infty$. As a
consequence, in order to study the meromorphic structure of the
zeta and eta functions, one can safely concentrate on the
asymptotic regime. In what follows, we show that the asymptotic
behavior of the true heat kernel ($K(z,z';t)$) coincides with the
asymptotic behavior of the parametrix constructed by pasting the
heat kernel in the cylinder, ($K_{c}(z,z';t)$), given by equation
(\ref{hk}), and the heat kernel in the double, which is a
boundaryless manifold.

Let $\rho (a,b)$ a $C^{\infty}$ function of the geodesic distance
from the boundary, $y$, such that $\rho(a,b)=0$ for $y\leq a$, and
$\rho (a,b)=1$ for $y\geq b$. Define the following smearing
functions

\beq
\begin{array}{cc}
\varphi_2=\rho(\frac14,\frac12)&\psi_2=\rho(\frac12,\frac34)\\
\varphi_1=1-\rho(\frac34,1)&\psi_1= 1-\psi_2\end{array}\,.\eeq

These smearing functions have the following characteristics:

1) In a collar neighborhood of the boundary, ${\mathcal
N}\times[0,1]$, they are functions of just the normal variable.

2) Each $\varphi_i$ is constant (has vanishing derivatives) and
equals 1 on the support of the corresponding $\psi_i$.

3) $\psi_1 +\psi_2=1$

We use these functions to construct a parametrix of the heat
kernel \beq Q(z,z';t)=\varphi_1(z)K_c
(z,z';t)\psi_1(z')+\varphi_2(z)K_d (z,z';t)\psi_2(z')\,,\eeq where
$K_c$ is the heat kernel in the infinite cylinder, given by
(\ref{hk}), and $K_d$ is the heat kernel in the double.

Now, at the level of operators, the error is given by

\beq \nn E(t)&=&K(t)-Q(t)=\int_0^t ds\frac{\partial}{\partial
s}(K(s)Q(t-s))\\ &=&\nn \int_0^t dsK(s) \left(-P^2_{\cal
B}+\frac{\partial}{\partial s}\right)Q(t-s)\\&=&\int_0^t ds
K(s)C(t-s)\,.\eeq

Thus, for the corresponding kernels one has \beq
K(z,z';t)-Q(z,z';t)=\int_0^t ds \int dw
K(z,w;s)C(w,z';t-s)\,,\label{inteq}\eeq where
\[C(w,z';t-s)=-\frac{d^2 \varphi_1}{dy^2}K_c(w,z';t-s)
\psi_1-\frac{d^2 \varphi_2}{dy^2}K_d(w,z';t-s)\psi_2-\] \beq
2\frac{d\varphi_1}{dy}\frac{dK_c}{dy}(w,z';t-s)
\psi_1-2\frac{d\varphi_2}{dy}\frac{dK_d}{dy}(w,z';t-s)\psi_2\,.\label{C}\eeq

As a consequence of the properties of the smearing functions (in
particular, due to 2)) $C(w,z';t)$ vanishes for $w\not\in
[\frac14,\frac34]\times {\mathcal N}$ and, in that region, it
vanishes when the distance $d(w,z')<\frac14$. In order to show
that $|C(w,z';t)|\leq c_1 e^{-\frac{c_2}{t}}$ for $d(w,z')\geq
\frac14$, we now prove the following

\bigskip

\noindent {\bf Proposition:} $K_c(z,z';t)$ is exponentially small
in $t$ as $t\rightarrow 0$ for $y\neq y'$. More precisely, it is
bounded by $Ct^{-1}\exp{\left(\frac{-(y-y')^2}{4t}\right)}$.

\bigskip

{\bf Proof}

As already shown, the projectors ${\cal P}_+$ and  ${\cal P}_-$
define two subspaces, where the heat kernel reduces, for each
$\omega$, to the fundamental solution of $\frac{\partial}{\partial
t} -\frac{\partial^2}{\partial y^2}+\omega^2$, with Dirichlet and
Robin boundary conditions respectively. The proposition is well
known to hold in the first case. In the second subspace, one has
\[ K_{c}(z,z';t)=\frac{1}{\beta(4\pi
t)^{\frac12}}\sum_{\omega}\phi_{\omega}^{\star}(x)
\phi_{\omega}(x') e^{-\omega^2
t}\left\{\left(e^{\frac{-\rho^2}{4t}}+e^{\frac{-\eta^2}{4t}}\right)\mathbf{1}\right.
\]
\beq \left.+\sqrt{(4\pi t)}\omega
\tanh{\Th}e^{u_{\omega}(\eta,t)^2}
erfc[u_{\omega}(\eta,t)]e^{\frac{-\eta^2}{4t}}\right\}
\label{hk2}\,.\eeq

So,\[|
K_{c}(z,z';t)|<\frac{1}{\beta}\sum_{\omega}|\phi_{\omega}^{\star}(x)
\phi_{\omega}(x')| e^{-\omega^2 t}\left[\frac{1}{\sqrt{\pi
t}}+\frac{2|\omega \tanh{(\Th)}|}{\sqrt{\pi
}}\right]e^{\frac{-\rho^2}{4t}}\,,\] where we have used that
$erfc(x)<\frac{2}{\sqrt{\pi}}e^{-x^2}$.

Now, using that $x\leq e^{\frac{x^2}{2}}$ and
$2|\phi_{\omega}^{\star}(x) \phi_{\omega}(x')|\leq
|\phi_{\omega}(x)|^2+|\phi_{\omega}(x')|^2$, we find
\[| K_{c}(z,z';t)|<\frac{1}{\beta 2\sqrt{(\pi
t)}}\sum_{\omega}\left(|\phi_{\omega}(x)|^2+|
\phi_{\omega}(x')|^2\right) e^{-\omega^2
t}\left[1+2e^{\frac{\omega^2\tanh^2{\Th}
t}{2}}\right]e^{\frac{-\rho^2}{4t}}\]
\[\leq\frac{3}{\beta 2\sqrt{(\pi
t)}}\sum_{\omega}\left(|\phi_{\omega}(x)|^2+|
\phi_{\omega}(x')|^2\right) e^{\frac{-\omega^2
t}{2}}e^{\frac{-\rho^2}{4t}}\,.\]

Finally, since the boundary is a boundaryless manifold, the
boundary heat kernel is bounded, on the diagonal of ${\mathcal
N}\times {\mathcal N}$, by $Ct^{-\frac{1}{2}}$, which leads us to
the announced result.

\bigskip

Now,  this estimate, and a similar one which is known to hold for
$K_d$, allow to show, from equation (\ref{C}), that
$|C(w,z';t)|\leq c_1 e^{-\frac{c_2}{t}}$. Thus, a convergent
series for the parametrix $Q$ can be constructed by iterating
(\ref{inteq}), and one can show that:

\bigskip

The exact heat kernel satisfies the bound
$|K(z,z';t)|<Ct^{-1}e^{-C'\frac{d^2 (z,z')}{t}}$ as $t\rightarrow
0$ and, on the diagonal, it differs from $Q$ according to
$|K(z,z;t)-Q(z,z;t)|\leq c e^{-\frac{c'}{t}}$ (for details see,
for instance, \cite{BW93}). So, one has, asymptotically

\beq Tr K\sim \int_0^1 \int_{{\mathcal N}} dy dx K_c\psi_1+
\int_{M}dy dx K_d\psi_2\,.\eeq

Observe that, due to the property 3) of the smearing functions,
the first term in $K_c$ (equation (\ref{hk})) adds up to the
contribution from the heat kernel in the double, to complete the
volume contribution to the trace. The remaining terms in the same
equation give the boundary contribution. Now, it is easy to show
that $\psi_1$ can be replaced by 1, and the integral $\int_0^1$
can be replaced by $\int_0^{\infty}$, since the difference
vanishes exponentially when $t\rightarrow 0$.

\end{appendix}

\bigskip

{\bf Acknowledgements:} Research of CGB and EMS was partially
supported by CONICET(PIP 0459/98) and UNLP(11/X298).

\end{document}